\documentstyle[12pt]{article}

\begin{document}

\begin{flushright}

IMSc/2018/02/02

\end{flushright} 

\vspace{2mm}

\vspace{2ex}

\begin{center}

{\large \bf Isotropic LQC and LQC--inspired Models} \\

\vspace{4ex}

{\large \bf with a massless scalar field} \\

\vspace{4ex}

{\large \bf as Generalised Brans--Dicke theories} \\

\vspace{8ex}

{\large  S. Kalyana Rama}

\vspace{3ex}

Institute of Mathematical Sciences, HBNI, C. I. T. Campus, 

\vspace{1ex}

Tharamani, CHENNAI 600 113, India. 

\vspace{2ex}

email: krama@imsc.res.in \\ 

\end{center}

\vspace{6ex}

\centerline{ABSTRACT}

\begin{quote} 

We explore whether generalised Brans -- Dicke theories, which
have a scalar field $\Phi$ and a function $\omega(\Phi)$, can be
the effective actions leading to the effective equations of
motion of the LQC and the LQC--inspired models, which have a
massless scalar field $\sigma$ and a function $f(m) \;$. We find
that this is possible for isotropic cosmology. We relate the
pairs $(\sigma, f)$ and $(\Phi, \omega)$ and, using examples,
illustrate these relations. We find that near the bounce of the
LQC evolutions for which $f(m) = sin \; m$, the corresponding
field $\Phi \to 0$ and the function $\omega(\Phi) \propto \Phi^2
\;$. We also find that the class of generalised Brans -- Dicke
theories, which we had found earlier to lead to non singular
isotropic evolutions, may be written as an LQC--inspired
model. The relations found here in the isotropic cases do not
apply to the anisotropic cases, which perhaps require more
general effective actions.

\end{quote}

\vspace{2ex}



\vspace{2ex}









\newpage

\vspace{4ex}

\begin{center}

{\bf 1. Introduction} 

\end{center}

\vspace{2ex}

In Loop quantum cosmology (LQC), the big bang singularities do
not arise \cite{ashtekar} -- \cite{status}. Starting with a
large $(3 + 1)$ dimensional universe and going back in time, one
finds that its physical volume decreases, reaches a non
vanishing minimum because of the quantum effects in LQC, and
starts increasing again. The densities and the temperatures of
the constituents of the universe do not diverge and remain
finite throughout. As explained in detail in the review
\cite{status}, these non singular evolutions arising due to
quantum effects can be described very well by effective
equations of motion. In the classical limit, these equations
reduce to Einstein's equations.

In a recent work \cite{k16}, we constructed the `LQC--inspired
models' by generalising empirically these effective LQC
equations to higher dimensions and to functions other than the
trigonometric ones. In \cite{k17}, we studied the cosmological
evolutions in these models using several examples.

It is natural now to search for an effective action which will
lead to the effective equations of the LQC--inspired models.
Besides being of interest for its own sake, such an action may
also be used practically : for example, one may use it to obtain
the generalisation of Schwarzschild solutions, or to study
spherically symmetric stars and their collapses. In the context
of LQC, $F(R)$ theories \cite{fr1, fr2, noo} have been used to
construct an effective action which leads to the isotropic
effective equations of LQC \cite{o08} -- \cite{o10}. It is shown
in \cite{o10} that the anisotropic case requires more general
$F(R, Q)$ theories where $Q = R_{\mu \nu} R^{\mu \nu} \;$.

It is very likely that the effective actions for the
LQC--inspired models may also be obtained using $F(R)$ theories
in the isotropic case or $F(R, Q)$ theories in the anisotropic
case. However, in this paper, we explore a different possibility
: the possibility of obtaining the effective actions of the LQC
and the LQC--inspired models using the $(d + 1)$ dimensional
scalar tensor theories \cite{will}. Our motivations for this
exploration are the following.

\begin{itemize}

\item

$F(R)$ theories can also be written as special cases of scalar
tensor theories with a particular value for the Brans -- Dicke
parameter $\omega$ and with a particular scalar field potential
which depends on the function $F$ \cite{fr1, fr2}.

\item

The effective actions will, clearly, be generalisations of the
standard Einstein's action and will reduce to it in the
classical limit. Now, the generalisations of Einstein's action
generically involve a scalar field which often appears as a
modulus field with no potential. For example: The Brans -- Dicke
theory arose as a theory which violates the strong, but obeys
the weak, equivalence principle of general relativity and, in
this theory, the gravitational constant was elevated to a
dynamical scalar field \cite{dicke}. In Kaluza -- Klein
theories, a scalar field appears upon dimensional reduction and
it describes the size of an extra dimension. Moduli fields are
ubiquitous in supergravity theories. In string theory effective
actions, a dilaton field is always present which is related to
the string coupling constant, see \cite{dp} in this context, or
equivalently to the size of the eleventh dimension in M theory.

\item

A massless scalar field is often used in LQC to define internal
time, with respect to which the quantum evolution of the
universe has a bounce, see \cite{status}. If this had been the
only way to define a time variable then one may say that a
scalar field with no potential appears in LQC also and plays the
role of internal time. However, as pointed out by the referee, a
time variable can be defined using pressureless dust or
radiation also \cite{husain, ppw}.


\item 

Long ago, in \cite{k95, bk}, we had found that non singular
evolution of a homogeneous isotropic universe is possible in
generalised Brans -- Dicke theories when a scalar field with no
potential is non minimally coupled with the coupling function
obeying certain conditions. It is possible that such theories
are related to some LQC--inspired models.
 
\end{itemize}

With these motivations in mind, we explore in this paper whether
$(d + 1)$ dimensional scalar tensor theories can be the
effective actions which will lead to the effective equations of
the LQC and the LQC--inspired models. We restrict ourselves to
models where there is only one scalar field and its potential
vanishes. And, on the other side, we restrict to a subclass of
scalar tensor theories, referred to as the generalised Brans --
Dicke theories, which contain only one scalar field, one
coupling function, and no scalar field potential.

The LQC and the LQC--inspired models have one scalar field
$\sigma \;$ with no potential and one function $f(m) \;$ where
$m$ is a certain Hubble-parameter-like variable in the
gravitational sector. The function $f(m) = sin \; m$ for LQC,
$f(m) = m$ gives Einstein's equations, and the classical limit
corresponds to taking the limit $m \to 0 \;$. One requires that
$f(m) \to m$ in the limit $m \to 0$ to ensure that the classical
limit leads to Einstein's equations.

The scalar field and the coupling function in the effective
action may be defined in several equivalent ways -- for example,
as the Brans -- Dicke field $\Phi$ and the function
$\omega(\Phi) \;$. In the original Brans -- Dicke theory,
$\omega$ is constant. Einstein theory follows in the limit
$\omega(\Phi) \to \infty \;$. Determining the pair $(\Phi, \;
\omega(\Phi))$ for a given $(\sigma, \; f(m))$ will then
specifiy the effective action corresponding to the effective
equations of motion of the LQC--inspired models.

In this paper, we consider isotropic effective equations and
find that one can indeed relate the pairs $(\sigma, \; f(m)) \;$
and $(\Phi, \; \omega(\Phi))$ -- actually, an equivalent pair
$(\phi, \; \Psi(\phi)) \;$. We describe the relations between
these pairs and they involve some integrations and functional
inversions. We can not carry out these operations explicitly for
any non trivial case. However, by studying a few explicit
examples and some limiting cases, we can illustrate several
important features of these relations.

\vspace{2ex}

The main results of this paper are the following.

\vspace{2ex}

We find that the LQC function $f(m) = sin \; m \;$
corresponds to an effective action where the function
$\omega(\Phi) \;$, equivalently $\Psi(\phi) \;$, has the
following properties: In the limit $m \to 0$ or $\pi \;$, the
function $\omega \to \infty$, equivalently $\phi \to \infty$ and
$\Psi \to const \;$. In the limit $m \to \frac {\pi} {2} \;$
from below, the field $\Phi \propto (\frac {\pi} {2} - m) \;$
and the function $\omega(\Phi) \propto \Phi^2 \;$, equivalently
$\phi \to - \infty$ and $\Psi \to \frac {\kappa \; \phi}
{\sqrt{d (d - 1)}} \;$. Another copy of $(\phi, \; \Psi(\phi))$
is needed when $m$ evolves further from $\frac {\pi} {2}$ to
$\pi \;$.

\vspace{2ex}

Also, we show that an LQC--inspired model which gives a non
singular evolution similar to that in LQC, but now with a much
flatter bounce, corresponds to a similar function $\omega(\Phi)$
as above, but now with $\omega(\Phi) \propto \Phi^{2 n} \;$ near
$\Phi \to 0 \;$ where the positive integer $n$ indicates how
flat the minimum is.

\vspace{2ex}

We also find that an LQC--inspired model with $f(m) \to m$ in
the limit $m \to 0$ and $f(m) \propto (m_s - m)^q$ in the limit
$m \to m_s$ from below, and with $\frac {1} {2} \le q < \frac
{d} {2 d - 1} \;$, corresponds to the generalised Brans -- Dicke
theories which led to non singular isotropic evolutions in our
earlier works \cite{k95, bk}.

\vspace{2ex}

Lastly, by studying an example, we find that the relations
obtained here between the pairs $(\sigma, \; f)$ and $(\phi, \;
\Psi)$ apply only to the homogeneous isotropic cases, and not to
the anisotropic ones. This means that anisotropic evolutions of
the LQC--inspired models can not be described by the scalar
tensor theories of the type considered here. A similar situation
arises in the LQC case also where the anisotropic evolution can
not be described by $F(R)$ theories and requires a further
generalisation to $F(R, Q)$ theories with $Q = R_{\mu \nu}
R^{\mu \nu} \;$ \cite{o10}. Perhaps a further generalisation is
needed here also, but its nature is not clear to us.

\vspace{2ex} 

The organisation of this paper is as follows. In section {\bf
2}, we write down the effective equations of motion of the
LQC--inspired models for the general anisotropic case. In
section {\bf 3}, we specialise these equations to the isotropic
case and to the scalar field with no potential. In section {\bf
4}, we briefly describe the relevant aspects of the generalised
Brans -- Dicke theory, write the action in several forms, write
the resulting equations of motion, and specialise them to the
anisotropic and the isotropic cases of interest here. In section
{\bf 5}, we relate the pairs $(\sigma, \; f)$ and $(\phi, \;
\Psi)$ and work out various examples and limiting cases. We
derive several results and also present a simple model for the
function $\Psi(\phi)$ which can capture several interesting
features of the evolution. In section {\bf 6}, we study an
example and find that the relations obtained in the isotropic
cases are not applicable for the anisotropic ones. In section
{\bf 7}, we conclude with a brief summary and some discussions.


\vspace{4ex}

\begin{center}

{\bf 2. Effective equations in the LQC--inspired models}

\end{center}

\vspace{2ex} 

Consider a $(d + 1)$ dimensional homogeneous, anisotropic
spacetime where $\; d \ge 3 \;$. Let the $d$ dimensional space
be toroidal, let $x^i$ be the coordinate and $L^i$ the
coordinate length of the $i^{th}$ direction, and let the line
element $d s$ be given by\footnote{ In the following, the
convention of summing over repeated indices is not always
applicable. Hence we will write explicitly the indices to be
summed over.}
\begin{equation}\label{ds}
d s^2 = - d t^2 + \sum_i a_i^2 \; (d x^i)^2
\end{equation}
where $\; i = 1, 2, \cdots, d \;$ and the scale factors $a_i$
are functions of $t$ only. We will write down the LQC--inspired
equations of motion for $a_i \;$ which are straightforward and
natural generalisations of the effective equations in LQC.


\vspace{2ex}

\begin{center}

{\bf LQC case}

\end{center}

\vspace{2ex} 

For the LQC case, $d = 3 \;$. In the Loop quantum gravity (LQG)
formalism, the canonical pairs of phase space variables consist
of an $SU(2)$ connection $A^i_a = \Gamma^i_a + \gamma K^i_a$ and
a triad $E^a_i$ of density weight one.\footnote{ See the review
\cite{status} for a complete description of the various LQG/C
terms and concepts mentioned here and in the following.} In
these expressions, $\Gamma^i_a$ is the spin connection defined
by the triad $e^a_i$, $\; K^i_a$ is related to the extrinsic
curvature, and $\gamma > 0$ and $\approx 0.2375$ is the
Barbero--Immirzi parameter of LQG, its numerical value being
suggested by the black hole entropy calculations. For the
anisotropic universe, whose line element $d s$ is given in
equation (\ref{ds}), one has $A^i_a \propto \hat{c}_i$ and
$E^a_i \propto \hat{p}_i$ where $\hat{c}_i \;$ will turn out to
be related to the time derivative of $a_i$, and $\hat{p}_i$ is
given by
\begin{equation}\label{pi}
\hat{p}_i = \frac {V} {a_i L_i} \; \; , \; \; \;
V = \prod_j {a_j L_j}
\end{equation}
with $V$ being the physical volume. The full expressions for
$A^i_a$ and $E^a_i$ contain various fiducial triads, cotriads,
and other elements, and are given in \cite{aw, status}. The non
vanishing Poisson brackets among $\hat{c}_i$ and $\hat{p}_j$ are
given by
\begin{equation}\label{cipi}
\{ \hat{c}_i, \; \hat{p}_j \} = \gamma \kappa^2 \;
\delta_{i j}
\end{equation}
where $\kappa^2 = 8 \pi G_{d +1}$ is the gravitational constant.
The effective equations of motion are given by the `Hamiltonian
constraint' ${\cal C}_H = 0$ and by the Poisson brackets of
$\hat{p}_i$ and $\hat{c}_i$ with ${\cal C}_H$ which give the
time evolutions of $\hat{c}_i$ and $\hat{p}_i \;$: namely, by
\begin{equation}\label{dynamics} {\cal C}_H = 0 \; \; , \; \; \;
(\hat{p}_i)_t = \{ \hat{p}_i, \; {\cal C}_H \} \; \; , \; \; \;
(\hat{c}_i)_t = \{ \hat{c}_i, \; {\cal C}_H \}
\end{equation}
where the $t-$subscripts denote derivatives with respect to $t
\;$. As reviewed in detail in \cite{status}, there exists an
effective ${\cal C}_H \;$ which leads to the equations of motion
which describe very well the quantum dynamics of LQC. In a
suitable limit, this effective ${\cal C}_H$ reduces to the
classical one which leads to Einstein's equations.

The expression for the ${\cal C}_H \;$ is of the form
\begin{equation}\label{chtot}
{\cal C}_H = H_{grav} (\hat{p}_i, \; \hat{c}_i) + H_{mat}
(\hat{p}_i \; ; \; \{ \phi_{mat} \}, \; \{ \pi_{mat} \}) 
\end{equation}
where $H_{grav}$ denotes the effective gravitational Hamiltonian
and $H_{mat}$ denotes a generalised matter Hamiltonian. In the
matter sector, the density $\tilde{\rho}$ and the pressure
$\tilde{p}_i$ in the $i^{th}$ direction are defined by
\begin{equation}\label{phati}
\tilde{\rho} = \frac {H_{mat}} {V} \; \; , \; \; \;
\tilde{p}_i = - \; \frac {a_i L_i} {V} \;
\frac {\partial H_{mat}} {\partial (a_i L_i)} \; \; .
\end{equation}
The pressure $\tilde{p}_i$ is thus, as to be physically
expected, proportional to the change in energy per fractional
change in the physical length in the $i^{th}$ direction. As
indicated in equation (\ref{chtot}), $\; H_{mat}$ is assumed to
be independent of $\hat{c}_i \;$. Since $\hat{c}_i$ will turn
out to be related to $(a_i)_t \;$, this assumption is equivalent
to assuming that matter fields couple to the metric fields but
not to the curvatures. It can also be shown \cite{k16} that,
irrespective of what $H_{grav}$ is, this assumption leads to the
standard conservation equation
\begin{equation}\label{ceqn}
\tilde{\rho}_t = \left( \frac {H_{mat}} {V} \right)_t
= - \; \sum_i (\tilde{\rho} + \tilde{p}_i) \;
\frac {(a_i)_t} {a_i} \; \; .
\end{equation}

In the gravitational sector, the effective $H_{grav}$ is given
by
\begin{equation}\label{hgravlqc}
H_{grav} = - \; \frac {V} {\gamma^2 \lambda_{qm}^2 \kappa^2}
\; \left( sin (\bar{\mu}^1 \hat{c}_1) \; sin (\bar{\mu}^2
\hat{c}_2) + \; cyclic \; \; terms \right)
\end{equation}
where $V = \sqrt{\hat{p}_1 \hat{p}_2 \hat{p}_3}$ is the physical
volume, $\lambda_{qm}^2 = \sqrt{ \frac {3} {4}} \; \gamma
\kappa^2$ is the quantum of area, and $\bar{\mu}^i = \frac
{\lambda_{qm} \hat{p}_i} {V}$ in what is referred to as the
$\bar{\mu}-$scheme. Classical $H_{grav}$ follows in the limit
$\bar{\mu}^i \hat{c}_i \to 0 \;$ where $sin \; (\bar{\mu}^i
\hat{c}_i) \to \bar{\mu}^i \hat{c}_i \;$.


\vspace{2ex}

\begin{center}

{\bf LQC--inspired models}

\end{center}

\vspace{2ex}

In recent papers \cite{k16, k17}, we have generalised the
effective LQC equations and studied their solutions. Our
generalisations are empirical but simple, straightforward, and
natural. We generalised from $(3 + 1)$ to $(d + 1)$ dimensions,
and generalised the trigonometric and the $\bar{\mu}$ functions
appearing in the effective $H_{grav}$ in equation
(\ref{hgravlqc}). In this paper, we will consider only the
generalisation of the trigonometric function, keeping the
$\bar{\mu}$ function the same as in the $\bar{\mu}-$scheme. Upon
generalisation :

\begin{itemize}

\item

The index $i$ takes the values $i = 1, 2, \cdots, d \;$ and $d
\ge 3 \;$ now.

\item

The canonical pairs of phase space variables are given by
$\hat{c}_i$ which will be related to $(a_i)_t$, and $\hat{p}_i$
which is given by equation (\ref{pi}). The non vanishing Poisson
brackets among $\hat{c}_i \;$ and $\hat{p}_j$ are given by
equation (\ref{cipi}) where $\gamma$ now characterises the
quantum of the $(d - 1)$ dimensional area given by
$\lambda_{qm}^{d - 1} \sim \gamma \kappa^2$ \cite{th1, th2, th3,
nb}.

\item

The effective equations of motion are given by equation
(\ref{dynamics}) where ${\cal C}_H \;$ is of the form given in
equation (\ref{chtot}). In the matter sector, the density
$\tilde{\rho}$ and the pressures $\tilde{p}_i$ are given by
equations (\ref{phati}), and they satisfy the standard
conservation equation (\ref{ceqn}).

\item

In the gravitational sector, the effective $H_{grav}$ in
equation (\ref{hgravlqc}) is now generalised to
\begin{equation}\label{hgrav}
H_{grav} = - \; \frac {V } {\gamma^2 \lambda_{qm}^2
\kappa^2} \; \sum_{i j \; (i < j)} f^i f^j
\end{equation}
where $V = \left( \prod_i \hat{p}_i \right)^{\frac{1}{d - 1}}$
is the $d$ dimensional physical volume and
\begin{equation}\label{fi}
f^i = f(m^i) \; \; , \; \; \;  
m^i = \bar{\mu}^i \; \hat{c}_i
\; \; , \; \; \;  
\bar{\mu}^i = \frac {\lambda_{qm} \; \hat{p}_i} {V} \; \; .
\end{equation}
The function $f(x)$ which appears in equation (\ref{fi}) is
arbitrary, but with the only requirement that $f(x) \to x$ as $x
\to 0 \;$ so that classical $H_{grav}$ is obtained in the limit
$m^i \to 0 \;$. It is easy to see that the LQC case follows upon
setting $d = 3$ and $f(x) = sin \; x \;$.

\end{itemize}

The equations of motion may now be obtained using the
generalised $H_{grav}$ given in equation (\ref{hgrav}). These
equations will describe the evolution of a $(d + 1)$ dimensional
homogeneous anisotropic universe in our LQC--inspired models.
The required algebra is straightforward but involved, and we
present only the final equations in a convenient form. Defining
$\lambda^i$, $\; \Lambda$, $\; G_{i j}$, and $G^{i j}$ by
\[
a_i = e^{\lambda^i} \; \; , \; \; \;
\Lambda = \sum_i \lambda^i \; \; , \; \; \;
G_{i j} = 1 - \delta_{i j} \; \; , \; \; \; 
G^{i j} = \frac{1}{d - 1} - \delta^{i j} \; \; , 
\]
the resulting equations of motion may be written as
\begin{eqnarray}
\sum_{i j} G_{i j} f^i f^j & = & 2 \; \gamma^2 \lambda_{qm}^2
\kappa^2 \; \tilde{\rho} \label{e1} \\
& & \nonumber \\
(m^i)_t + \sum_j \frac {(m^i - m^j) \; X_j} {(d - 1) \; \gamma
\lambda_{qm}} & = & - \; \gamma \lambda_{qm} \kappa^2 \; \sum_j
G^{i j} \; (\tilde{\rho} + \tilde{p}_j) \label{e2} \\
& & \nonumber \\
(\gamma \lambda_{qm}) \; \lambda^i_t & = &
\sum_j G^{i j} X_j \label{e3} \\
& & \nonumber \\
\tilde{\rho}_t & = & - \; \sum_i (\tilde{\rho} + \tilde{p}_i) \;
\lambda^i_t \label{e4} 
\end{eqnarray}
where $X_i$ is given by 
\[
X_i = g_i \sum_j G_{i j} f^j \; \; , \; \; \;
g_i = \frac{d f(m^i)} {d m^i} \; \; . 
\] 

Einstein's equations follow when $f(m^i) = m^i \;$. Then $g_i =
1$, equation (\ref{e3}) gives $(\gamma \lambda_{qm}) \;
\lambda^i_t = m^i \;$, and, after a little algebra, equations
(\ref{e1}) and (\ref{e2}) give the Einstein's equations for a
$(d + 1)$ dimensional homogeneous anisotropic universe :
\begin{eqnarray}
\sum_{i j} G_{i j} \; \lambda^i_t \; \lambda^j_t
& = & 2 \kappa^2 \; \tilde{\rho} \label{ee1} \\
& & \nonumber \\
\lambda^i_{t t} + \Lambda_t \;
\lambda^i_t & = & \kappa^2 \; \sum_j G^{i j} \;
(\tilde{\rho} - \tilde{p}_j) \; \; . \label{ee3}
\end{eqnarray}


\vspace{4ex}

\begin{center}

{\bf 3. Isotropic case with a massless scalar field}

\end{center}

\vspace{2ex}

For the isotropic case, one has $(\tilde{p}_i , \; m^i , \; f^i
, \; a_i ) = (\tilde{p} , \; m , \; f , \; a ) \;$. Then
\begin{equation}\label{hiso}
\lambda^i_t = \frac {a_t} {a} \equiv h \; \; , \; \; \;
X_i = (d - 1) \; g f \; \; , \; \; \; 
g_i = g = \frac{d f} {d m} \; \; . 
\end{equation}
Also $e^{\Lambda} = a^d$ and $\Lambda_t = d \; h \;$. Equations
of motion (\ref{e1}) -- (\ref{e4}) now give
\begin{eqnarray}
f^2 & = & \frac {2 \; \gamma^2 \lambda_{qm}^2 \kappa^2 \;
\tilde{\rho}} {d \; (d - 1)} \label{eiso1} \\
& & \nonumber \\
m_t & = & - \; \frac {\gamma \lambda_{qm} \kappa^2} {d - 1} \;
(\tilde{\rho} + \tilde{p})
\label{eiso2} \\
& & \nonumber \\
h & = & \frac {a_t} {a} \; = \; \frac {g \; f}
{\gamma \lambda_{qm}} \label{eiso3} \\
& & \nonumber \\
\tilde{\rho}_t & = & - \; d \; h \; (\tilde{\rho} + \tilde{p})
\; \; . \label{eiso4} 
\end{eqnarray}

Consider a massless scalar field $\sigma$ with no potential. It
is straightforward to show that its density, pressures, and the
equation of motion in a homogeneous, isotropic universe are
given by
\begin{equation}\label{tilderhopi}
\tilde{\rho} = \tilde{p} = \frac {(\sigma_t)^2} {2}
\; \; , \; \; \;
\sigma_{t t} + d \; h \; \sigma_t = 0 \; \; . 
\end{equation}
Equation (\ref{eiso4}) is then satisfied. Equations
(\ref{eiso1}) and (\ref{eiso2}) now become
\begin{eqnarray}
f^2 & = & \frac {\gamma^2 \lambda_{qm}^2 \kappa^2}
{d \; (d - 1)} \; (\sigma_t)^2 \label{tseiso1} \\
& & \nonumber \\
m_t & = & - \; \frac {\gamma \lambda_{qm} \kappa^2}
{d - 1} \; (\sigma_t)^2 \; \; . \label{tseiso2} \\
& & \nonumber 
\end{eqnarray}
With no loss of generality, let $\sigma_t = \sqrt{2
\tilde{\rho}} = \frac {\sqrt{d ( d - 1)}} {\gamma \lambda_{qm}
\kappa} f \;$ where the square roots are always taken with a
positive sign.  Then equations (\ref{tilderhopi}) --
(\ref{tseiso2}) give
\begin{equation}\label{fa}
\frac {\sigma_t} {\sigma_{t0}} =
\sqrt{ \frac {\tilde{\rho}} {\tilde{\rho}_0} } =
\frac {f} {f_0} = \left( \frac {a_0} {a} \right)^d 
\end{equation}
where the $0-$subscripts denote the values at an initial time
$t_0 \;$ and
\begin{equation}\label{dmdsigma} 
\frac {d m} {f^2} = - \; c_{qm} \; d t \; \; , \; \; \;
\frac {d m} {f} = - \; c_1 \; d \sigma
\end{equation}
where $c_{qm} = \frac{d} {\gamma \lambda_{qm}}$ and $c_1 =
\sqrt{ \frac {\kappa^2 \; d} {d - 1} } \;$. Thus, for a given
function $f(m) \;$, equations (\ref{fa}) and (\ref{dmdsigma})
give $a(m)$, $\; t(m)$, and $\sigma(m) \;$ which, in principle,
then give the solutions $m(t)$, $\; a(t)$, and $\sigma(t) \;$.

For example, let $f(m) = m$ which leads to Einstein's equations.
Then, using $d h_0 = c_{qm} m_0$ and after a little algebra, one
obtains that
\begin{equation}\label{f=msoln}
\frac {f_0} {f} = \frac {m_0} {m} =
\left( \frac {a} {a_0} \right)^d =
e^{c_1 \; (\sigma - \sigma_0)} =
d h_0 \; \left( t - t_0 + \frac {1} {d h_0} \right) \; \; .
\end{equation}
It can also be shown that the evolution near any simple zero of
$f$ is same as that given by Einstein's equations upto a
constant scaling of $t \;$.

As another example, consider the evolution near a maximum of $f
\;$. Let $f$ reach a maximum at $m = m_b$ and, near $m_b \;$,
let
\begin{equation}\label{fmax}
f(m) \; \simeq \; f_b \;
\left( 1 - f_1 (m_b - m)^{2 n} \right)
\end{equation} 
where $f_b$ and $f_1$ are positive constants and $n$ is a
positive integer which indicates how flat the maximum is. Let
$t_b$ be the time when $f$ reaches its maximum. Then, as $t \to
t_b \;$, it follows from equations (\ref{dmdsigma}) that
\begin{eqnarray}
m_b - m & \simeq & f_b^2 \; c_{qm} (t - t_b) \nonumber \\
& & \nonumber \\
c_1 (\sigma - \sigma_b) & \simeq &
f_b \; c_{qm} (t - t_b) \nonumber \\ 
& & \nonumber \\
a & \simeq & a_{mn} \; \left( 1 + a_1 (t - t_b)^{2 n} \right)
\label{fmxsoln} 
\end{eqnarray}
where $\sigma_b$ is a constant, $a_{mn} = a_0 \; \left( \frac
{f_0} {f_b} \right)^{ \frac {1} {d} }$, and $a_1 = \frac {f_1}
{d} \; (f_b^2 c_{qm})^{2 n} \;$. If $f(m) = sin \; m \;$ then
$m_b = \frac {\pi} {2}$ and $f_b = 2 f_1 = n = 1 \;$ in equation
(\ref{fmax}). It can then be checked that equations
(\ref{fmxsoln}) are consistent with the explicit solution given,
for example, in \cite{k17}.


\vspace{4ex}

\begin{center}

{\bf 4.  Generalised Brans--Dicke theory} 

\end{center}

\vspace{2ex} 

The effective equations (\ref{tilderhopi}) -- (\ref{tseiso2})
or, equivalently, (\ref{fa}) and (\ref{dmdsigma}) describe the
evolution of a $(d + 1)$ dimensional homogeneous isotropic
universe with a massless scalar field in the LQC--inspired
models. For such an universe, the line element $d s$ and the
metric $g_{\mu \nu}, \; \mu, \nu = 0, 1, 2, \cdots, d \;$, are
given by
\begin{equation}\label{dsiso}
d s^2 = \sum_{\mu \nu} g_{\mu \nu} \; dx^\mu d x^\nu
= - d t^2 + a^2 \sum_i (d x^i)^2 
\end{equation}
where the scale factor $a$ is a function of $t$ only.  It is
natural to search for an effective action which, if exists, will
lead to covariant equations that generalise the standard
Einstein's equations and, for the above line element, lead to
the effective equations (\ref{fa}) and (\ref{dmdsigma}).

We assume that such an effective action exists and proceed to
construct it invoking the following line of reasoning. The
action should contain the metric field $g_{\mu \nu}$ and a
scalar field $\phi$ which, in general, may be different from
$\sigma \;$. The construction should also involve an arbitrary
function to act as an equivalent for the function $f$ of the
LQC--inspired models. The required action cannot be the
canonical minimally coupled action for $\phi$ with a potential
$V(\phi)$ since it cannot lead to a non singular evolution of
$a(t)$ with a bounce.

However, the scalar field $\phi$ may couple non minimally. Then,
irrespective of whether there is a potential for the scalar
field, the general non minimal coupling will involve a function
which may act as an equivalent for the function $f \;$. Non
minimally coupled scalar field appears naturally, and
generically with no potential, in several contexts. For example,
it appears as a dilaton field in string theories and as a moduli
field in supergravity theories. It also appears in Brans--Dicke
theories where the gravitational constant is elevated to a
dynamical scalar field. Also, in our past works in this context
\cite{k95, bk}, we had found that non singular evolution of a
homogeneous isotropic universe is possible for appropriate
choices of the non minimal coupling function. For these reasons,
and also due to the motivations listed in the Introduction, we
consider in this paper the generalised Brans--Dicke theories
where a scalar field with no potential is non minimally coupled
with a non trivial coupling function. \footnote{ \label{f(r)} In
LQC, $(3 + 1)$ dimensional effective actions have been
constructed in \cite{o08} -- \cite{o10} by generalising
Einstein's term $R$ to $F(R)$ and finding an appropriate
function $F$ which will give the isotropic LQC evolution. Any
$F(R)-$theory, including that for LQC, can be written as a
scalar -- tensor theory with the Brans--Dicke constant $\omega =
0$ and with a potential that depends on $F \;$ \cite{fr1, fr2}.
Similar approach may also work for the present $(d + 1)$
dimensional LQC--inspired models for any arbitrary function $f
\;$, but we will not pursue it in this paper.}


\vspace{4ex}

\begin{center}

{\bf  Action : different forms}

\end{center}

\vspace{2ex}

The $(d + 1)$ dimensional action for a scalar field $\phi$ with
no potential and with a non minimal coupling function may be
written in several equivalent forms. Consider the action
$S_{st}$ for a scalar -- tensor theory given by
\begin{equation}\label{sst}
S_{st} = \frac {1} {2 \kappa^2} \; \int \; d^{d + 1} x \;
\sqrt{- g} \; \left( \Omega \; R - B \; (\nabla \phi)^2 \right)
\end{equation}
where $\Omega$ and $B$ are functions of $\phi \;$ and the scalar
potential $V(\phi) = 0 \;$. There is a freedom in defining the
scalar field which may be used to specify the functions $\Omega$
and $B$ in a convenient form. For example, setting $\phi =
\Omega = \Phi$ and $B = \frac {\omega(\Phi)} {\Phi}$ gives the
generalised Brans -- Dicke action
\begin{equation}\label{sbd}
S_{bd} \; = \; \frac {1} {2 \kappa^2} \; \int d^{d + 1} x \;
\sqrt{ - g} \; \left( \Phi \; R - \frac {\omega(\Phi)} {\Phi}
\; (\nabla \Phi)^2 \right)
\end{equation}
which elevates the gravitational constant to a spacetime
dependent dynamical scalar field $\Phi \;$ and the Brans --
Dicke constant $\omega$ to a function of $\Phi \;$ now. Or, the
action $S_{st}$ may instead be written as
\begin{equation}\label{sPsi}
S_\Psi \; = \; \int d^{d + 1} x \; \sqrt{ - g} \; \;
e^{ (d - 1) \Psi } \; \left( \frac{R} {2 \kappa^2}
- \frac { {\cal A} } {2} \; (\nabla \phi)^2 \right)
\end{equation}
where we have set 
\begin{equation}\label{calA}
B = \kappa^2 \; \Omega \; {\cal A} \; \; , \; \; \;
\Omega = e^{(d - 1) \; \Psi} \; \; , \; \; \; 
{\cal A} = 1 - \frac {d (d - 1)} {\kappa^2} \; (\Psi_\phi)^2
\end{equation}
and the $\phi-$subscripts denote derivatives with respect to
$\phi \;$. The function $\Psi(\phi)$ is now the non minimal
coupling function. The action $S_\Psi$ follows upon setting
$g_{* \mu \nu} = e^{2 \Psi} g_{\mu \nu} $ in the `Einstein
frame' action $S_*$ given by
\begin{equation}\label{s*}
S_* = \int d^{d + 1} x \; \sqrt{ - g_*} \; \left( \frac{R_*}
{2 \kappa^2} - \frac {(\nabla_* \phi)^2} {2} \right)
\end{equation}
where the action for $g_{* \mu \nu}$ is as in Einstein's theory,
the scalar field $\phi$ has a canonical kinetic term, and $\phi$
is coupled minimally to $g_{* \mu \nu} \;$. However, other
fields and the probes in the theory are assumed to couple
minimally, not to $g_{* \mu \nu} \;$, but to $g_{\mu \nu} \;$.
Hence, in the Einstein frame, they experience a force due to
scalar field $\phi$ and will not fall freely along the geodesics
of $g_{* \mu \nu} \;$.

Note that $(\Phi , \; \omega(\Phi))$ in the action $S_{bd}$ and
$(\phi, \; \Psi(\phi))$ in the action $S_\Psi$ may be related to
each other easily. If given $\phi$ and $\Psi(\phi)$ then it
follows from equations (\ref{sbd}) and (\ref{sPsi}) that
\begin{equation}\label{omega}
\Phi = e^{(d - 1) \; \Psi} \; \; , \; \; \; 
\omega = \frac {\kappa^2 \; {\cal A}}
{(d - 1)^2 \; (\Psi_\phi)^2}
= \frac {\kappa^2} { (d - 1)^2 \; (\Psi_\phi)^2}
\; - \; \frac {d} {d - 1} \; \; . 
\end{equation}
Inverting the first expression gives, in principle, $\phi(\Phi)$
which then gives $\omega(\Phi) \;$. Thus, the limit $\Psi(\phi)
\to const$ which leads to Einstein's theory can now be seen
easily to correspond to the limit $\omega \to \infty \;$. If
given $\Phi$ and $\omega(\Phi)$ instead, it then follows from
equation (\ref{omega}) that
\begin{equation}\label{Psi}
(d - 1) \; \Psi = ln \; \Phi \; \; , \; \; \; 
\kappa \; d \phi = \frac {d \Phi} {\Phi} \; 
\left( \omega + \frac {d} {d - 1} \right)^{\frac {1} {2}} 
\; \; . 
\end{equation}
The function $\phi(\Phi) \;$ now follows upon an integration.
Inverting this function then gives, in principle, $\Phi(\phi)$
and thereby $\Psi(\phi) \;$.


\vspace{4ex}

\begin{center}

{\bf Equations of motion from $\mathbf S_{st}$}

\end{center}

\vspace{2ex} 

\noindent {\bf General :} 
Consider the action $S_{st}$ given in equation (\ref{sst}).
Defining $t_{\mu \nu}$ by
\[
\kappa^2 \; t_{\mu \nu} \; = \; B \left( \nabla_\mu \phi
\nabla_\nu \phi - \frac {g_{\mu \nu}} {2} \; (\nabla \phi)^2
\right) \; \; , 
\]
the equations of motion following from the action $S_{st}$ may
be written as
\begin{equation}\label{est1} 
\Omega \; \left (R_{\mu \nu} - \frac {g_{\mu \nu}} {2} \; R
\right) \; = \; \kappa^2 \; t_{\mu \nu} + \nabla_\mu \nabla_\nu
\Omega - g_{\mu \nu} \; \nabla^2 \Omega
\end{equation}
and 
\begin{equation}\label{est2} 
2 B \; \nabla^2 \phi + \; B_\phi \; (\nabla \phi)^2
+ \Omega_\phi \; R \; = \; 0 \; \; .
\end{equation}
It follows from equation (\ref{est1}) that
\[
\Omega \; R \; = \; \frac {2 d} {d - 1} \; \nabla^2 \Omega
+ B \; (\nabla \phi)^2
\]
and then from equation (\ref{est2}) that 
\begin{equation}\label{est4}
\left(B + \frac {d \; (\Omega_\phi)^2} {(d - 1) \; \Omega}
\right) \; \nabla^2 \phi + \left( \frac {B_\phi} {2}
+ \frac {B \; \Omega_\phi} {2 \; \Omega}
+ \frac {d \; \Omega_\phi \; \Omega_{\phi \phi}}
{(d - 1) \; \Omega} \right) \; (\nabla \phi)^2 \; = \; 0 \; \; .
\end{equation}
For the generalised Brans -- Dicke action $S_{bd} \;$ given in
equation (\ref{sbd}), we have $\phi = \Omega = \Phi$ and $B =
\frac {\omega(\Phi)} {\Phi} \;$. Equation (\ref{est4}) then
simplifies to
\begin{equation}\label{est5}
2 \; \left( \omega + \frac {d} {d - 1} \right) \; \nabla^2 \Phi
+ \omega_\Phi \; (\nabla \Phi)^2 \; = \; 0 \; \; . 
\end{equation}

\vspace{4ex}

\noindent {\bf Anisotropic case :} 
Let the metric $g_{\mu \nu}$ be given by
\begin{equation}\label{dsst}
d s^2 = \sum_{\mu \nu} g_{\mu \nu} \; dx^\mu d x^\nu
= - d t^2 + \sum_i a_i^2 \; (d x^i)^2 
\end{equation}
where the scale factors $a_i$ and the scalar field $\phi$ depend
on $t$ only. Defining $a_i = e^{\lambda^i}$ and $\Lambda =
\sum_i \lambda^i \;$, the non vanishing components of $R^\mu_{\;
\; \nu}$ and $\nabla^\mu \nabla_\nu \; \Omega \;$, and thereby
$R$ and $\nabla^2 \Omega \;$, are given by
\begin{eqnarray*}
R^t_{\; \; t} \; = \; \Lambda_{t t}
+ \sum_i \left( \lambda^i_t \right)^2 & , & 
R^i_{\; \; i} \; = \; \lambda^i_{t t}
+ \Lambda_t \; \lambda^i_t \\
& & \\
\nabla^t \nabla_t \; \Omega \; = \; - \; \Omega_{t t}
& , & 
\nabla^i \nabla_i \; \Omega \; = \; - \; \Omega_t \lambda^i_t \\
& & \\
R \; = \; 2 \; \Lambda_{t t} + (\Lambda_t)^2
+ \sum_i \left( \lambda^i_t \right)^2 & , & \nabla^2 \Omega
\; = \; - \; \Omega_{t t} - \Omega_t \Lambda_t \; \; .
\end{eqnarray*}
It then follows from equations (\ref{est1}) and (\ref{est2})
that
\begin{eqnarray}
\Omega \; \left( (\Lambda_t)^2 - \sum_i (\lambda^i_t)^2 \right)
& = & B \; (\phi_t)^2 - 2 \Lambda_t \; \Omega_t \label{est21} \\
& & \nonumber \\
\Omega \; \left( \lambda^i_{t t} + \Lambda_t \lambda^i_t \right)
+ \Omega_t \; \lambda^i_t & = & - \; \frac { \Omega_{t t}
+ \Lambda_t \; \Omega_t} {d - 1} \label{est22} \\
& & \nonumber \\
2 B \; \left( \phi_{t t} + \Lambda_t \; \phi_t \right) & = &
- \; B_\phi \; (\phi_t)^2 + \Omega_\phi \; R \; \; .
\label{est23}
\end{eqnarray}

\vspace{4ex}

\noindent {\bf Isotropic case :}
For the isotropic case, $a_i = a$, $\; \lambda^i_t = \frac {a_t}
{a} = h$, and $\Lambda_t = d h \;$. Equations (\ref{est21}) and
(\ref{est22}) then give
\begin{eqnarray}
d (d - 1) \; \Omega \; h^2 & = & B \; (\phi_t)^2
- 2 d \; h \; \Omega_t \label{est31} \\
& & \nonumber \\
- \; (d - 1) \; \Omega \; h_t & = & B \; (\phi_t)^2
+ \Omega_{t t} - h \; \Omega_t \; \; . \label{est32}
\end{eqnarray}


\vspace{4ex}

\begin{center}

{\bf Equations of motion from $\mathbf S_*$ and their solutions}

\end{center}

\vspace{2ex} 

\noindent {\bf Anisotropic case :} 
Let the metric $g_{* \mu \nu}$ in the Einstein frame be given by
\begin{equation}\label{ds*}
d s_*^2 = \sum_{\mu \nu} g_{* \mu \nu} \; dx^\mu d x^\nu
= - d T^2 + \sum_i A_i^2 \; (d x^i)^2 
\end{equation}
where the scale factors $A_i$ and the scalar field $\phi$ depend
on $T$ only. The equations of motion resulting from $S_*$ follow
from equations (\ref{est21}) -- (\ref{est23}) upon setting
$\Omega = 1$ and $B = \kappa^2 \;$. Defining $A_i = e^{l^i}$ and
$L = \sum_i l^i \;$, we obtain
\begin{eqnarray}
& & (L_T)^2 - \sum_i (l^i_T)^2 \; = \; \kappa^2 \; (\phi_T)^2
\label{est*1} \\
& & \nonumber \\
& & l^i_{T T} + L_T \; l^i_T \; = \; 
\phi_{T T} + L_T \; \phi_T \; = \; 0 \; \; . \label{est*3}
\end{eqnarray}
These equations can be solved explicitly. Equations
(\ref{est*3}) give
\begin{equation}\label{liT}
l^i_T = l^i_{T0} \; e^{L_0 - L}
\; \; , \; \; \;
\phi_T = \phi_{T0} \; e^{L_0 - L} 
\end{equation}
where the $0-$subscripts denote the values at an initial time
$T_0 \;$. We then have $L_T = L_{T0} \; e^{L_0 - L}$ where
$L_{T0} = \sum_i l^i_{T0} \;$ and, hence,
\begin{equation}\label{e^L}
e^{L - L_0} = L_{T0} \; \tilde{T} \; \; , \; \; \;
\tilde{T} = T - T_0 + \frac {1} {L_{T0}} \; \; .
\end{equation}
Writing $l^i_{T0} = \alpha^i \; L_{T0} \;$ and $\kappa \;
\phi_{T0} = \beta \; L_{T0} \;$, we obtain $\sum_i \alpha^i =
1$,
\begin{equation}\label{anisoln}
e^{l^i - l^i_0} = \left( L_{T0} \; \tilde{T} \right)^{\alpha^i}
\; \; , \; \; \; e^{\kappa (\phi - \phi_0)}
= \left( L_{T0} \; \tilde{T} \right)^\beta \; \; ,
\end{equation}
and then, from quation (\ref{est*1}), the constraint $\sum_i
(\alpha^i)^2 + \beta^2 = 1 \;$.

\vspace{4ex}

\noindent {\bf Isotropic case :}
For the isotropic case, we have $e^{l^i} = A$, $\; l^i_T = \frac
{A_T} {A} = H$, and $L_T = d H \; $. Hence,
\begin{equation}\label{est*21} 
d (d - 1) \; H^2 \; = \; - \; (d - 1) \; H_T \; = \;
\kappa^2 \; (\phi_T)^2 \; \; . 
\end{equation}
Also, $\alpha^i = \frac {1} {d}$ and $\beta = \sqrt{ \frac {d -
1} {d} } \;$ in equations (\ref{anisoln}). Hence $\frac {\kappa}
{\beta} = \sqrt{ \frac {\kappa^2 \; d} {d - 1} } = c_1 \;$ and
the isotropic solutions may be written as 
\begin{equation}\label{isoln}
\left( \frac {A} {A_0} \right)^d \; = \;
e^{c_1 \; (\phi - \phi_0)} \; = \; L_{T 0} \; \tilde{T}
\; = \; \left( \frac {\phi_{T0}} {\phi_T} \right) \; \; . 
\end{equation}


\vspace{4ex}

\begin{center}

{\bf Equations of motion from $\mathbf S_\Psi$}

\end{center}

\vspace{2ex} 

The equations of motion resulting from $S_\Psi$ follow similarly
from equations (\ref{est21}) -- (\ref{est23}) upon using
equation (\ref{calA}) for the functions $\Omega$ and $B \;$.
They may also be obtained by noting that $S_*$ gives $S_\Psi$
upon setting $g_{* \mu \nu} = e^{2 \Psi} g_{\mu \nu} \;$.
Equations (\ref{dsst}) and (\ref{ds*}) then give
\begin{equation}\label{tTA}
d T \; = e^\Psi \; d t \; \; , \; \; \;
A_i \; = e^\Psi \; a_i \; \; \longrightarrow \; \; \;
l^i \; = \Psi + \lambda^i \; \; .
\end{equation}
Substituting these expressions in equations (\ref{est*1}) and
(\ref{est*3}) must then give the equations obtained by
substituting equations (\ref{calA}) for $\Omega$ and $B$ in
equations (\ref{est21}) -- (\ref{est23}). We have verified that
this is indeed the case.


\vspace{4ex}

\begin{center}

{\bf 5. Relating the functions $f$ and $\Psi \;$} 

\end{center}

\vspace{2ex}

Consider a $(d + 1)$ dimensional homogeneous isotropic universe
with a massless scalar field. In an LQC or LQC--inspired model
which is specified by a function $f$, the evolution is described
by equations (\ref{fa}) and (\ref{dmdsigma}), rewritten below
for ease of reference :
\begin{equation}\label{5fa} 
\frac {f} {f_0} = \left( \frac {a_0} {a} \right)^d
\; \; , \; \; \; 
\frac {d m} {f^2} = - \; c_{qm} \; d t \; \; , \; \; \; 
\frac {d m} {f} = - \; c_1 \; d \sigma 
\end{equation}
where $c_{qm} = \frac{d} {\gamma \lambda_{qm}}$ and $c_1 =
\sqrt{ \frac {\kappa^2 \; d} {d - 1} } \;$.  These equations
give $a(m), \; t(m)$, and $\sigma(m)$ for a given function $f(m)
\;$. In generalised Brans -- Dicke theory, the evolution is
described by equations (\ref{isoln}) and (\ref{tTA}) with $A_i =
A$ and $a_i = a$, rewritten below for ease of reference :
\begin{equation}\label{5isoln}
\left( \frac {\phi_T} {\phi_{T_0}} \right) \; = \;
\left( \frac {A_0} {A} \right)^d \; = \;
e^{- \; c_1 \; (\phi - \phi_0)} 
\end{equation}
which give $T(\phi)$ and $A(\phi) \;$, and 
\begin{equation}\label{5tTA}
d T \; = e^\Psi \; d t \; \; , \; \; \;
A \; = e^\Psi \; a 
\end{equation}
which then give $t(\phi)$ and $a(\phi) \;$ for a given function
$\Psi(\phi) \;$. The functions $f$ and $\Psi$ can then be
related to each other. In the following, we take the initial
values $f_0, \; \sigma_{t0}$, and $\phi_{T0}$ to be positive for
the sake of definiteness, and set $\sigma_0 = \phi_0 = 0$ with
no loss of generality.

We now study the relation between the functions $f$ and $\Psi
\;$. The scalar field $\sigma$ may be different from $\phi$ and,
hence, the relation between them also needs to be studied. Now,
after some algebra and with $c_m = c_1 c_\sigma f_0$ and
$c_\sigma = \frac {\sigma_{t0}} {e^{\Psi_0} \; \phi_{T0}} \;$,
it follows from equations (\ref{5fa}) -- (\ref{5tTA}) that
\begin{eqnarray}
f & = & f_0 \; e^{d \; (\Psi - \Psi_0)
\; - \; c_1 \; \phi} \label{fphi} \\
& & \nonumber \\
d m & = & - \; c_m \; \; e^{(2 d - 1) \; (\Psi - \Psi_0)
\; - \; c_1 \; \phi} \; \; d \phi \; \; . \label{mphi} \\
& & \nonumber \\
d \sigma & = & c_\sigma \; \; e^{(d - 1) \; (\Psi - \Psi_0)}
\; \; d \phi \label{sigmaphi}
\end{eqnarray}
It then follows from these equations, or from $\frac {d t} {a} =
\frac {d T} {A} \;$, that
\begin{equation}\label{dtaTA}
\left( \frac {f_0} {f} \right)^{2 - \frac {1} {d}} d m \; = \;
- \; c_m \; e^{ \left(1 - \frac {1} {d} \right) \; c_1 \; \phi}
\; \; d \phi \; \; . 
\end{equation}

Consider equations (\ref{fphi}) -- (\ref{dtaTA}). If a function
$\Psi(\phi)$ is given then equation (\ref{fphi}) gives $f(\phi)
\;$; equation (\ref{mphi}) gives $m(\phi) \;$; and equation
(\ref{sigmaphi}) gives $\sigma(\phi) \;$. The function $f(m)$
follows now, in principle, from $f(\phi)$ and $m(\phi) \;$.
Thus, a given function $\Psi(\phi)$ determines the functions
$f(m) \;$ and $\sigma(\phi) \;$.

If a function $f(m)$ is given then $\sigma(m) \;$ follows from
the last equation in (\ref{5fa}) or from equations (\ref{fphi})
-- (\ref{sigmaphi}). Equation (\ref{dtaTA}) gives $\phi(m) \;$
and, then, equation (\ref{fphi}) gives $\Psi(m) \;$. The
function $\Psi(\phi)$ follows now, in principle, from $\phi(m)$
and $\Psi(m) \;$; and the function $\phi(\sigma)$ from
$\sigma(m)$ and $\phi(m) \;$. Thus, a given function $f(m) \;$
determines the functions $\Psi(\phi)$ and $\phi(\sigma) \;$.


\vspace{4ex}

\begin{center}

{\bf Examples and Limiting cases}

\end{center} 

\vspace{2ex}

The functions $f(m)$ and $\Psi(\phi)$ can be related to each
other as described above. This will then relate the
corresponding LQC--inspired model and the generalised
Brans--Dicke theory. However, we are not able to carry out
explicitly all the required integrations and functional
inversions for any non trivial function, including the LQC
function $f(m) = sin \; m \;$. Nevertheless, several important
features of the relation between these models can be understood
by studying a few explicit examples and some limiting cases.


\vspace{4ex}

\begin{center}

{\bf Example (1) : $\mathbf f(m) = m$ } 

\end{center}

\vspace{2ex}

Consider the example $f(m) = m \;$. Then the last equation in
(\ref{5fa}) gives $m = m_0 \; e^{- c_1 \sigma} \;$.  Equation
(\ref{dtaTA}) then gives $m = m_0 \; c_\sigma^{ \frac {d} {1 -
d}} \; e^{- c_1 \phi} \;$. Equation (\ref{fphi}) then gives
$c_\sigma \; e^{(d - 1) \; (\Psi - \Psi_0)} = 1$ which implies
that $\Psi = const$ and, hence, that $\Psi = \Psi_0$ and
$c_\sigma = 1 \;$ which, in turn, gives $\phi = \sigma \;$.
Note that equations (\ref{fphi}) and (\ref{mphi}) give $g =
\frac {d f} {d m} = \left( \frac {c_1 - d \Psi_\phi} {c_1
c_\sigma} \right) e^{- (d - 1) (\Psi - \Psi_0)}$ from which also
it follows that $c_\sigma = 1$ when $f(m) = m$ and $\Psi =
\Psi_0 \;$.

Thus, when $f(m) = m$ one has $\Psi = const$ and $\phi = \sigma
\;$. This, of course, corresponds to Einstein's theory. It is
easy to see that a similar result follows for any linear
function $f = \alpha_1 m + \alpha_2$ where $\alpha_1$ and
$\alpha_2$ are constants. Indeed, now start with $\Psi (\phi) =
const = \Psi_0 \;$. Then equation (\ref{fphi}) gives $f = f_0 \;
e^{- c_1 \; \phi} \;$; equation (\ref{sigmaphi}) gives $\sigma =
c_\sigma \; \phi \;$; and, equation (\ref{mphi}) gives $m = m_0
+ c_\sigma \; (f - f_0) \;$ which is of the form $f = \alpha_1 m
+ \alpha_2 \;$.


\vspace{4ex}

\begin{center}

{\bf Example (2) : $\mathbf \Psi(\phi) = k \; c_1 \phi + \Psi_0$}

\end{center}

\vspace{2ex}

Consider the example where $\Psi(\phi) = k \; c_1 \phi + \Psi_0
\;$, and $k$ and $\Psi_0$ are constants. It follows from
equation (\ref{omega}) that this corresponds to a Brans -- Dicke
theory with
\begin{equation}\label{komega} 
\Phi = \Phi_0 \; e^{k (d - 1) \; c_1 \phi} \; \; , \; \; \;
\omega = \frac {1 - k^2 d^2} {d ( d - 1) \; k^2} 
\end{equation}
and $\Phi_0 = e^{(d - 1) \Psi_0} \;$. For this example,
equations (\ref{fphi}) -- (\ref{mphi}) give
\begin{eqnarray}
f & = & f_0 \; e^{(k d - 1) \; c_1 \phi} \label{kf} \\
& & \nonumber \\
c_1 \; \sigma & = & \frac { c_\sigma } {k (d - 1) } \;
\left( e^{k (d - 1) \; c_1 \phi} - 1 \right) \label{kphi} \\
& & \nonumber \\
m - m_0 & = & m_1 \; \left( 1 - e^{{\cal D} \; c_1 \phi}
\right) \label{km} \\
& & \nonumber \\
\Longrightarrow \; \; \;
f(m) & = & f_0 \; \left( \frac {m_1 + m_0 - m} {m_1} \right)^q
\label{kfm} 
\end{eqnarray}
where
\[
m_1 = \frac {c_m} {c_1 \; {\cal D}} \; \; , \; \; \; 
{\cal D} = k (2 d - 1) - 1 \; \; , \; \; \;
q = \frac {k d - 1} {{\cal D}} \; \; .
\]
Let $k > \frac {1} {d} \;$ so that ${\cal D}$ and $q$ are
positive. Then, in the limit $m \to m_s$ where $m_s = m_0 + m_1
\;$, the scalar field $\phi \to - \infty$, the function $f \to
0$, and the scale factor $a \propto f^{- \frac {1} {d}} \to
\infty \;$. Also, equations (\ref{5isoln}) and (\ref{5tTA}) give
$t(\phi)$ :
\begin{equation}\label{kt}
c_t \; (t - t_0) \; = \; e^{(1 - k) \; c_1 \phi} - 1
\; \; , \; \; \; 
c_t = (1 - k) \; c_1 \phi_{t0} 
\end{equation}
for $k \ne 1 \;$, and $\phi_{t0} \; (t - t_0) = \phi \;$ for $k
= 1 \;$. Hence, in the limit $\phi \to - \infty \;$, the time $t
\to t_s = t_0 - \frac {1} {c_t}$ if $k < 1$ and $t \to - \infty$
if $k \ge 1 \;$. 

It now follows that if $k \ge 1 \;$ then $q \ge \frac {1} {2} $
and, in the limit $m \to m_s$, the evolution is non singular
because the scale factor $a \to \infty \;$ and the time $t \to -
\infty \;$. This is the same result obtained in \cite{k17} in
the context of LQC--inspired models, and in \cite{k95, bk} in
the context of generalised Brans -- Dicke theories. Also, if $k
= 1$ then it is easy to see that the entire evolution is non
singular and the scale factor $a(t)$ evolves exponentially in
time : As $\phi$ varies from $- \infty$ to $+ \infty \;$, the
time $t \propto \phi$ also varies the same way, $m$ varies from
$m_s$ to $- \infty \;$, the function $f(t) \propto e^{(*) \; t}$
varies from $0$ to $\infty$, and the scale factor $a(t)\propto
e^{- (*) \; t}$ varies from $\infty$ to $0 \;$, the $(*)$s being
some positive constants.

Note that if $\frac {1} {d} < k < \infty$ then $0 < q < \frac
{d} {2 d - 1} \;$. This upper bound on $q$ may also be
understood as follows. Integrating equation (\ref{dtaTA}), which
follows from $\frac {d t} {a} = \frac {d T} {A} \;$, gives
\[
- \; \int^m_{m_0} d m' \; f^{\frac {1} {d} - 2} \; = \;
(*) \; \int^\phi_{\phi_0} d \phi' \;
e^{ \frac {d - 1} {d} \; c_1 \; \phi'}
\]
where $(*)$ is an unimportant constant. Now, the $\phi-$integral
in the above equation is finite in the limit $\phi \to - \infty
\;$. In the corresponding limit, $m \to m_s \;$ and $f \sim
(m_s - m)^q \;$. Therefore the $m-$integral in the above
equation will also be finite in this limit only if $q < \frac
{d} {2 d - 1} \;$.


\vspace{4ex}

\begin{center}

{\bf Example (3) : $\mathbf f(m) \; \simeq \; f_b \;
\left( 1 - f_1 (m_b - m)^{2 n} \right)$}

\end{center}

\vspace{2ex}

Consider the example where $f(m)$ has a maximum at $m_b$ and,
near its maximum, is given by $f(m) \; \simeq \; f_b \; \left( 1
- f_1 \; \tilde{m}^{2 n} \right)$ where $f_b$ and $f_1$ are
positive constants, $\tilde{m} = (m_b - m)$, and $n$ is a
positive integer which indicates how flat the maximum is. As $f$
reaches its maximum and then decreases, the scale factor $a$
reaches a minimum and increases again -- it has a `bounce'. The
scalar field $\phi$ decreases as $m$ increases, see equation
(\ref{mphi}). In the limit $m \to m_b$ from below then two cases
are possible: {\bf (i)} $\phi \to \phi_b$, a finite value, or
{\bf (ii)} $\phi \to - \infty \;$.

\vspace{2ex}

\noindent
{\bf Case (i) $\mathbf \phi \to \phi_b \;$:}
Let $\phi$ have a finite value $\phi_b$ at the bounce. Let
$\tilde{\phi} = \phi - \phi_b$ and $\tilde{\Psi} = \Psi - \Psi_b
\;$. Here and in the following, we encounter various unimportant
constants. We will denote them all by $(*)$, keeping only their
signs. It then follows from equations (\ref{dtaTA}) and
(\ref{fphi}) that, near the bounce,
\begin{eqnarray}
\tilde{m} & \simeq & (*) \; \tilde{\phi} \label{b*dtaTA} \\
& & \nonumber \\
e^{\tilde{\Psi}} & \simeq & e^{ \frac {c_1} {d} \; \tilde{\phi}}
\; (1 - (*) \; \tilde{m}^{2 n}) \label{b*fphi} \\
& & \nonumber \\
\Longrightarrow \; \; \; 
\Psi_\phi & \simeq & \frac {c_1} {d}
\; - \; (*) \; \tilde{m}^{2 n - 1} \; \; . \label{b*Psiphi}
\end{eqnarray}
It is now illuminating to change over from the present $\phi$
and $\Psi(\phi)$ to the field $\Phi$ and the function
$\omega(\Phi)$ of the generalised Brans -- Dicke theory, given
by equations (\ref{omega}). Let $\Phi_b = e^{(d - 1) \; \Psi_b}$
and $\tilde{\Phi} = \Phi - \Phi_b \;$. Using the above
expressions, it then follows that
\begin{eqnarray}
\Phi & = & e^{(d - 1) \; \Psi} \; \simeq \; \Phi_b \; \left( 1
+ (*) \; \tilde{m} \right)  \; \; , \nonumber \\
& & \nonumber \\
\omega(\Phi) & \simeq & (*) \; \tilde{m}^{2 n - 1} \; \simeq \;
(*) \; (\Phi - \Phi_b)^{2 n - 1} \; \; . \label{b*omega}
\end{eqnarray}

\vspace{2ex}

\noindent
{\bf Case (ii) $\mathbf \phi \to - \infty \;$:}
Let $\phi$ diverge to $- \infty$ at the bounce. It then follows
from equations (\ref{dtaTA}) and (\ref{fphi}) that
\begin{eqnarray}
\tilde{m} & \simeq & (*) \; e^{ \left(1 - \frac {1} {d} \right)
\; c_1 \; \phi} \label{*dtaTA} \\
& & \nonumber \\
e^\Psi & \simeq & (*) \; e^{ \frac {c_1} {d} \; \phi}
\; (1 - (*) \; \tilde{m}^{2 n}) \label{*fphi} \\
& & \nonumber \\
\Longrightarrow \; \; \; 
\Psi_\phi & \simeq & \frac {c_1} {d}
\; - \; (*) \; \tilde{m}^{2 n} \; \; . \label{*Psiphi}
\end{eqnarray}
Changing over to the field $\Phi$ and the function
$\omega(\Phi)$ using equations (\ref{omega}) and the above
expressions, it then follows that
\begin{eqnarray}
\Phi & = & e^{(d - 1) \; \Psi} \; \simeq \; (*) \; \tilde{m}
\; \; , \nonumber \\
& & \nonumber \\
\omega(\Phi) & \simeq & (*) \; \tilde{m}^{2 n}
\; \simeq \; (*) \; \Phi^{2 n} \; \; . \label{*omega} 
\end{eqnarray}

These expressions for $\Phi$ and $\omega(\Phi)$ illuminate
nicely the salient features of the relation between the
LQC--inspired models and the corresponding generalised Brans --
Dicke theories. When the function $f$ reaches its maximum,
equivalently the scale factor $a$ reaches its minimum, the field
$\Phi$ may reach a finite value or may vanish; and the function
$\omega(\Phi)$ will have a zero of odd or even order
respectively. The positive integer $n$, which indicates how flat
the maximum is, corresponds to the order of the zero of $\omega
\;$: $\omega(\Phi) \propto (\Phi - \Phi_b)^{2 n - 1} \;$ has a
zero of odd order if $\Phi$ reaches a finite value at the
bounce; and $\omega(\Phi) \propto \Phi^{2 n} \;$ has a zero of
even order if $\Phi$ vanishes at the bounce.

Also, it follows that $\tilde{m}(t) \simeq (*) \; (t - t_b)$
where $t_b$ is the time when $m = m_b \;$, see the second
equation in (\ref{5fa}). Now, as $m$ approaches $m_b$ from below
and evolves past it, $\tilde{m}$ crosses zero and becomes
negative. The corresponding evolution in case {\bf (i)} is
straightforward. Consider case {\bf (ii)}. The scalar field
$\phi$, being $\propto (ln \; \tilde{m})$, is not well defined
when $\tilde{m}$ becomes negative. Another copy of $\phi$ and
the function $\Psi$ seems to be needed to dictate further
evolution. However, the Brans -- Dicke field $\Phi \propto
\tilde{m}$ and can smoothly cross zero and become negative.
Further evolution will depend on the function $\omega \;$. It
can be seen from the action $S_{bd}$ given in equation
(\ref{sbd}) that, during this crossing, the kinetic terms for
both the metric and the scalar field change signs. The
significance of these changes of signs is not clear to us,
perhaps they are unphysical. However, the equations of motion
(\ref{est5}) and (\ref{est21}) -- (\ref{est32}) remain the same,
and we have verified that the corresponding solutions continue
smoothly across $t_b$ where $\Phi(t)$ crosses zero and
$\omega(\Phi) \propto \Phi^{2 n} \;$.


\vspace{4ex}

\begin{center}

{\bf Example (4) : $\mathbf f(m) = m$ near $\mathbf 0 \; ; \; \;
\; \mathbf f_{(2)}$ or $\mathbf f_{(3)}$ near $\mathbf m_r$}

\end{center}

\vspace{2ex}

Consider the example where $f(m) = m$ near $m = 0 \;$; remains
positive with all its derivatives finite in the interval $0 < m
< m_r \;$; and is given by $f_{(2)} \propto (m_r - m)^q$ as in
Example {\bf (2)}, or by $f_{(3)} \propto \left( 1 - (*) \; (m_r
- m)^{2 n} \right)$ as in Example {\bf (3)}, in the limit $m \to
m_r \;$.

The evolution corresponding to such a function $f$ is as
follows, see \cite{k17}. As $m$ increases from $0$, the
evolution is initially as in Einstein's theory : the time $t$
decreases from $\infty$ and the scale factor $a$ decreases from
$\infty \;$. The evolution proceeds smoothly until $m \to m_r
\;$. In the limit $m \to m_r$ if the function $f \to f_{(2)} \to
0 \;$ then the scale factor $a \to \infty \;$, the time $t \to
t_s$ or $- \infty$ depending on whether $0 < 2 q < 1$ or $2 q
\ge 1 \;$, and the evolution will be singular or non singular
respectively. If the function $f \to f_{(3)} \;$ in the limit $m
\to m_r$ then the scale factor $a$ will reach a minimum and will
have a bounce. Further evolution requires specification of
$f(m)$ beyond $m_r$ which will then be related to another copy
of $\phi$ and another function $\Psi \;$. Choosing an $f$
symmetric around its maximum, {\em e.g.} $f(m) = sin \; m \;$,
will require the same $\Psi \;$ and will lead to a symmetric
bounce.

Consider now the function $\Psi(\phi)$ which may correspond to
the function $f(m)$ of the present Example {\bf (4)}.  We assume
that this function $\Psi(\phi)$ will be such that $\Psi \to
const$ as $\phi \to \infty$; $\; \Psi \to k \; c_1 \phi$ as
$\phi \to - \infty$; and all the derivatives of $\Psi$ are
finite for all $\phi \;$. A simple model for such a function
$\Psi(\phi)$ is given by
\begin{equation}\label{spsi}
e^{- s \Psi} \; = \; c + b \; e^{ - s k \; c_1 \phi}
\; \; \; \longleftrightarrow \; \; \; 
e^{s \Psi} \; = \; \frac {e^{s k \; c_1 \phi}}
{b + c \; e^{s k \; c_1 \phi}}
\end{equation}
where $s, \; k, \; b$, and $c$ are positive constants. Using
\[
\Psi_\phi = k \; c_1 \left( 1 - c \; e^{s \Psi} \right)
= \frac {k \; c_1 b} {b + c \; e^{s k \; c_1 \phi}} 
\]
and equations (\ref{omega}), it follows that the Brans -- Dicke
field $\Phi$ and the function $\omega$ corresponding to the
present model are given by
\begin{eqnarray} 
\Phi^s & = & \frac {e^{(d - 1) s k \; c_1 \phi}}
{\left( b + c \; e^{s k \; c_1 \phi} \right)^{d - 1} }
\nonumber \\
& & \nonumber \\
\omega & = & \frac { \left(b + c \; e^{s k \; c_1 \phi}
\right)^2 \; - \; b^2 k^2 d^2} { d (d - 1) \; b^2 k^2 }
\; \; . \label{sphiomega}
\end{eqnarray}
The scalar field $\phi \;$, and then the function $\omega \;$,
can be expressed in terms of $\Phi$ but this is not necessary
here.  By construction, we have $\Psi \to k \; c_1 \phi$ as
$\phi \to - \infty \;$. It therefore follows that if $k > \frac
{1} {d}$ then, in the limit $\phi \to - \infty \;$, one obtains
the function $f(m) \propto (m_r - m)^q$ in the limit $m \to m_r$
and with $q > 0 \;$. Note that $q$ and $k$ are related by $q =
\frac {k d - 1} {k (2 d - 1) - 1}$ and, hence, that if $\frac
{1} {d} < k < 1$ then $0 < 2 q < 1 \;$ and if $1 \le k < \infty$
then $1 \le 2 q < \frac {2 d} {d - 1} \;$.

On the other hand, the choice $k = \frac {1} {d}$ and $s = 2 n
(d - 1)$ will give the function $f(m)$ as in Example {\bf (3)}.
With $k = \frac {1} {d}$ and in the limit $\phi \to - \infty
\;$, one has
\[
\Phi \; \simeq \; (*) \; e^{(d - 1) k \; c_1 \phi}
\; \; , \; \; \;
\omega \; \simeq \; (*) \; e^{s k \; c_1 \phi}
\; \simeq \; (*) \; \Phi^{\frac {s} {d - 1}} \; \; .
\]
Choosing $s = 2 n (d - 1)$ will now give a $\omega(\Phi)$ having
a zero of order $2 n$ at $\Phi = 0$ and, hence, will lead to a
function $f(m)$ which has a maximum of the type considered in
Example {\bf (3)}. This can be seen more explicitly also. Note
that equation (\ref{fphi}) gives
\begin{equation}\label{sa}
\left( \frac {f} {f_0} \right)^s \; = \; \left( \frac {a} {a_0}
\right)^{- s d} \; = \;
\left( \frac {b + c} {b + c \; e^{s k \; c_1 \phi}} \right)^d
\; e^{(kd - 1) s \; c_1 \phi} \; \; .
\end{equation}
We are not able to do the integrations needed to obtain
$t(\phi)$ explicitly. Hence, consider the limit $\phi \to -
\infty \;$ where $\Psi \to k \; c_1 \phi \;$. For $k = \frac {1}
{d} \;$, equations (\ref{5fa}) and (\ref{mphi}) now give
\[
(m_r - m) \; \simeq \; (*) \; (t - t_r)
\; \simeq \; (*) \; e^{\frac {d - 1} {d} \; c_1 \phi}
\]
where $t_r$ is a constant. If $s = 2 n \; (d - 1)$ then $e^{s k
\; c_1 \phi} \propto (t - t_r)^{2 n} \;$ and equations
(\ref{sa}) give $f(m)$ and $a(t)$ as in Example {\bf (3)} :
\[
f(m) \; \propto \;
\left( 1 - (*) \; (m_r - m)^{2 n} \right)
\; \; , \; \; \; a(t) \; \propto \;
\left( 1 + (*) \; (t - t_r)^{2 n} \right) \; \; .
\]

Further evolution beyond $m_r$ requires specification of $f(m)$
beyond $m_r \;$. For example, let the function be symmteric
around $m_r$, namely let $f(m) = f(2 m_r - m)$ for $m_r \le m
\le 2 m_r \;$. It is then easy to see that as $m$ varies from
$m_r$ to $2 m_r$, the evolution will be described by the same
function $\Psi(\phi)$, now with $\phi$ varying from $- \infty$
to $+ \infty \;$.

Consider now the LQC function $f(m) = sin \; m \;$ which is
symmetric around $\frac {\pi} {2}$ and for which $n = 1 \;$. Let
$\Psi_{lqc}(\phi)$ be the corresponding function in the
generalised Brans -- Dicke theory. We are not able to obtain
$\Psi_{lqc}(\phi)$ in an explicit form. However, it follows from
Examples {\bf (1)} and {\bf (3)} that $\phi \to + \infty$ and
$\Psi_{lqc}(\phi) \to const$ in the limit $m \to 0 \;$; that
$\phi \to - \infty$ and $\Psi_{lqc}(\phi) \to \frac {c_1 \phi}
{d}$ in the limit $m \to \frac {\pi} {2}$ from below; and that
the field $\Phi \propto (\frac {\pi} {2} - m) \;$ and the
function $\omega(\Phi) \propto \Phi^2 \;$ in the limit $m \to
\frac {\pi} {2} \;$. Furthermore, as $m$ varies from $\frac
{\pi} {2}$ to $\pi$, it also follows that the evolution will be
described by the same function $\Psi_{lqc}(\phi)$, now with
$\phi$ varying from $- \infty$ to $+ \infty \;$.


\vspace{4ex}

\begin{center}

{\bf 6. Anisotropic case} 

\end{center}

\vspace{2ex}

Consider the example of the LQC--inspired models where $f(m) =
m$ near $m = 0 \;$; remains positive with all its derivatives
finite in the interval $0 < m < m_r \;$; and $f(m) \propto (m_r
- m)^q$ in the limit $m \to m_r \;$. Consider the anisotropic
case. The line element $d s$ is now given by equation (\ref{ds})
and the equations of motion by (\ref{e1}) -- (\ref{e4}) with
$\tilde{\rho} = \tilde{p}_i = \frac {(\sigma_t)^2} {2} \;$. In a
recent work \cite{k17}, we have analysed the cosmological
evolution in such LQC--inspired models and have shown that if $2
q > 1$ then the corresponding anisotropic evolution is non
singular.  The analysis is straightforward but involved and,
hence, will not be presented here.

The function $\Psi(\phi)$ corresponding to such an $f(m)$ is
given in Example {\bf (4)} in the isotropic case. If the
LQC--inspired model for a given function $f(m)$ and the
generalised Brans -- Dicke theory with the corresponding
function $\Psi(\phi)$ are equivalent to each other then this
equivalence may be obtained by studying the isotropic case, and
it should be applicable for the anisotropic case also. It turns
out that this is not the case.

To see the absence of this equivalence in the anisotropic case,
consider now the generalised Brans -- Dicke theories in the
limit $\phi \to - \infty \;$. In this limit, let $\Psi \to
\tilde{k} \; (\kappa \phi)$ where $\tilde{k} > 0$ is a
constant. The solutions for time $t$ and the scale factors $a_i$
can now be obtained in this limit : Equations (\ref{anisoln})
give the solutions for the scale factors $A_i$ and the field
$\phi$ in Einstein frame; equations (\ref{tTA}) then give the
time $t$ and the scale factors $a_i \;$. We rewrite these
expressions below for ease of reference :
\begin{eqnarray}
A_i = A_{i0} \; \left( L_{T0} \; \tilde{T} \right)^{\alpha^i}
& , &
e^{\kappa \phi} = \left( L_{T0} \; \tilde{T} \right)^\beta
\nonumber \\ 
& & \nonumber \\
d t = e^{ - \Psi} \; d T & , & a_i = e^{ - \Psi} \; A_i
\label{anidtdT} 
\end{eqnarray}
where $\sum_i \alpha^i = \sum_i (\alpha^i)^2 + \beta^2 = 1$ and
$\tilde{T} = T - T_0 + \frac {1} {L_{T0}} \;$. In the following,
we take $\beta > 0 \;$ with no loss of generality. This implies,
since $\beta$ is non vanishing, that $\vert \alpha^i \vert < 1
\;$.

\vspace{2ex}

Consider now the limit $\tilde{T} \to 0$ where $\phi \to -
\infty \;$. In this limit, let $\Psi \to \tilde{k} \; (\kappa
\phi)$ where $\tilde{k} > 0$ is a constant. Then $e^\Psi \propto
\tilde{T}^{\tilde{k} \beta} \to 0 \;$ and equations
(\ref{anidtdT}) give
\begin{equation}\label{taik}
t \; \simeq \; \frac {(*) \; \tilde{T}^{1 - \tilde{k} \beta}}
{1 - \tilde{k} \beta} \; \; , \; \; \;
a_i \; \simeq \; (*) \; \tilde{T}^{\alpha^i - \tilde{k} \beta}
\end{equation}
if $\tilde{k} \beta \ne 1 \;$, and $t \simeq (ln \; \tilde{T})$
if $\tilde{k} \beta = 1 \;$. Since $\vert \alpha^i \vert < 1
\;$, it now follows that if $\tilde{k} \beta \ge 1 \;$ then
$(\alpha^i - \tilde{k} \beta) < 0$ for all $i \;$ and,
therefore, the evolution is non singular because all the scale
factors $a^i \to \infty \;$ and the time $t \to - \infty \;$.
Also, it follows from equations (\ref{omega}) that $\omega$ is
given in this limit by
\begin{equation}\label{8omega} 
\omega = \frac {1} { (d - 1)^2 \; \tilde{k}^2}
\; - \; \frac {d} {d - 1} \; \; . 
\end{equation}

Note that $\beta$ is assumed to be positive and non vanishing,
but it can be very small : $0 < \beta \ll 1 \;$. This means that
if the anisotropic evolution must be non singular for all such
values of $\beta$ also then $\tilde{k} \beta > 1 \;$ for all
$\beta \ll 1$ also and, hence, it follows that $\tilde{k} \gg 1
\;$. In turn, this implies that the corresponding $\omega$ must
be close to $- \frac {d} {d - 1} \;$. Conversely, if $\tilde{k}$
is large but finite then the anisotropic evolution of $a^i(t)$
will not always be non singular. There will be set of values $\{
\alpha^i \} \;$ of non zero measure for which $\beta = \sqrt{ 1
- \sum_i (\alpha^i)^2} \; < \; \frac {1} {\tilde{k}}$ and,
hence, the corresponding anisotropic evolution will be singular.

In the isotropic case, $\omega = - \; \frac {d} {d - 1} \;$
corresponds to the function $f(m) \propto (m_s - m)^q$ where $q
= \frac {d} {2 d - 1} \;$. However, in the LQC--inspired models,
a non singular anisotropic evolution is possible for any value
of $q > \frac {1} {2} \;$, see \cite{k17}. It therefore follows
that the relations between the functions $\Psi$ and $f$ studied
here apply only to the homogeneous isotropic cases, and not to
the anisotropic ones.

In this context, note that a similar situation occurs also in
the case of $F(R)$ theories which are constructed to give the
isotropic LQC evolution. As shown in \cite{o10}, to be able to
describe the aniostropic evolution also, one needs to generalise
$F(R)$ theories to $F(R, Q)$ theories where $Q = R_{\mu \nu}
R^{\mu \nu} \;$. Perhaps then it is not surprising that the
generalised Brans -- Dicke theories, which are constructed here
to give the isotropic evolution of the LQC--inspired models, are
not able to describe the anisotropic evolution also. Moreover,
it is likely that some further generalisation is needed but the
nature of such a generalistion is not clear to us.


\vspace{4ex}

\begin{center}

{\bf 7. Conclusion}

\end{center}

\vspace{2ex}

We first give a brief summary. In this paper, we explore whether
a subclass of scalar tensor theories can be the effective
actions which will lead to the effective equations of motion of
the LQC and the LQC--inspired models. We consider models where
there is only one scalar field with no potential, and consider
the generalised Brans -- Dicke theories which contain only one
scalar field, one coupling function, and no scalar field
potential. Thus, a scalar field $\sigma$ and a function $f(m)$
of the LQC--inspired models need to be related to the field
$\Phi$ and a function $\omega(\Phi)$ of the corresponding
action.

We consider the isotropic case and find the relation between
these two pairs. We can not do explicit calculations for non
trivial cases. Hence, we study a few explicit examples and some
limiting cases and, using them, illustrate several important
features of these relations. For example, we find that near the
bounce of the LQC evolutions for which $f(m) = sin \; m \;$, the
corresponding field $\Phi \to 0$ and the function $\omega(\Phi)
\propto \Phi^2 \;$. Also, we find in this paper that the class
of generalised Brans -- Dicke theories, which was found in our
earlier works to lead to non singular isotropic evolutions, may
be written as an LQC--inspired model with an appropriate
function $f(m) \;$.

We further find that the relations between the LQC-inspired
models and the generalised Brans -- Dicke theories do not apply
to the anisotropic cases. A similar situation arises in LQC
where also the anisotropic cases can not be described by $F(R)$
theories, and a further generalisation to $F(R, R_{\mu \nu}
R^{\mu \nu})$ theories is needed. Perhaps a further
generalisation is needed here also, but the nature of such a
generalisation is not clear to us.

We now conclude by mentioning a few topics for further studies.

It is desireable to have an unique effective action which will
give the effective equations of motion of the LQC and the
LQC--inspired models. Such an action should give the equations
of motion, for example, of the LQC of Bianchi type I, II, IX
models. We are not aware of a physical principle that guarantees
the existence of such an action and, furthermore, its
uniqueness. It seems miraculous in this context that even the
effective equations of motion exist, and that they describe well
the quantum effects of the LQC in various Bianchi type models.

The effective equations of motion themselves are very useful.
They can be generalised empirically, as we have done recently to
obtain the LQC--inspired models, and can be used efficiently as
a laboratory to model and study a variety of cosmological
evolutions -- four or higher dimensional, isotropic or
anisotropic, with or without compactifications, et cetera.

An effective action, if it exists and can be found, will be even
more useful. It may be used, for example, to study spherical
stars and their collapses, or to study how the perturbations of
a homogeneous universe evolve when the universe undergoes a
bounce. It is, of course, too much to expect that such an action
will describe all the quantum effects of LQG itself in these new
situations. Nevertheless, it may be expected to lead to new
effects. We note here that the class of generalised Brans --
Dicke theories, which was found in our earlier works to lead to
non singular isotropic evolutions, also leads to interesting new
effects when applied to stars \cite{kg, bk}.

A class of theories, called mimetic gravity and degenerate
higher order scalar tensor theories, have been proposed which
contain higher order derivative terms but without the associated
pathologies. Also, in \cite{cm}, a class of mimetic theories has
been constructed which resolve the cosmological singularities.
These theories also lead to the effective isotropic equations of
LQC, as shown in \cite{bss, llnw}. See \cite{blln} also where
mimetic theories are used to construct non singular black hole
solutions. It is thus of interest to explore the connection, if
any, between such theories and generalised Brans -- Dicke
theories considered here.


\vspace{4ex}

{\bf Acknowledgement:} We thank the referee for her/his
suggestions and for pointing out the references \cite{husain,
ppw, bss, llnw}.


\end{document}